\documentclass[%
 reprint, superscriptaddress,
 amsmath,amssymb,
 aps,
prb,
]{revtex4-1}

\usepackage{graphicx}
\usepackage{dcolumn}
\usepackage{bm}
\usepackage{physics}
\usepackage{verbatim}
\usepackage{color}
\usepackage{hyperref}
\usepackage{float}

\begin{document}

\title{A Spin Entanglement Witness for Quantum Gravity}
\author{Sougato Bose}
\affiliation{Department of Physics and Astronomy, University College London, Gower Street, WC1E 6BT London, UK}

\author{Anupam Mazumdar}
\affiliation{Van Swinderen Institute
University of Groningen
9747 AG Groningen, 
The Netherlands}

\author{Gavin W. Morley}
\affiliation{Department of Physics, University of Warwick, Gibbet Hill Road, Coventry CV4 7AL, UK}

\author{Hendrik Ulbricht}
\affiliation{Department of Physics and Astronomy, University of Southampton, SO17 1BJ, Southampton, UK}

\author{Marko Toro\v{s}}
\affiliation{Department of Physics and Astronomy, University of Southampton, SO17 1BJ, Southampton, UK}

\author{Mauro Paternostro}
\affiliation{CTAMOP, School of Mathematics and Physics, Queen's University Belfast, BT7 1NN Belfast, UK}

\author{Andrew Geraci}
\affiliation{Department of Physics, University of Nevada, Reno, USA, 89557}

\author{Peter Barker}
\affiliation{Department of Physics and Astronomy, University College London, Gower Street, WC1E 6BT London, UK}

\author{M. S. Kim}
\affiliation{QOLS, Blackett Laboratory, Imperial College, London SW7 2AZ, UK}

\author{Gerard Milburn}
\affiliation{QOLS, Blackett Laboratory, Imperial College, London SW7 2AZ, UK}
\affiliation{Centre for Engineered Quantum Systems,
School of Mathematics and Physics,
The University of Queensland, QLD 4072 Australia.
}

\begin{abstract}
{Understanding gravity in the framework of quantum mechanics is one of the great challenges in modern physics. Along this line, a prime question is to find whether gravity is a quantum entity subject to the rules of quantum mechanics. It is fair to say that there are no feasible ideas yet to test the quantum coherent behaviour of gravity directly in a laboratory experiment. Here, we introduce an idea for such a test based on the principle that two objects cannot be entangled without a quantum mediator.  We show that despite the weakness of gravity, the phase evolution induced by the gravitational interaction of two micron size test masses in adjacent matter-wave interferometers can detectably entangle them even when they are placed far apart enough to keep Casimir-Polder forces at bay. We provide a prescription for witnessing this entanglement, which certifies gravity as a quantum coherent mediator, through simple correlation measurements between two spins: one embedded in each test mass. Fundamentally, the
above entanglement is shown to certify the presence of non-zero off-diagonal terms in the coherent state basis of the gravitational field modes.}
\end{abstract}

\maketitle


Quantizing gravity is one of the most intensively pursued areas of physics \cite{Oriti,Kiefer}. However, the lack of empirical evidence for quantum aspects of gravity has lead to a debate on whether gravity is a quantum entity. This debate includes a significant community who subscribe to the breakdown of quantum mechanics itself at scales macroscopic enough to produce prominent gravitational effects \cite{Penrose,Diosi,GRW-review,Hendrik-Gravity}, so that gravity need not be a quantized field in the usual sense. Indeed it is quite possible to treat gravity as a classical agent at the cost of including additional stochastic noise \cite{Milburn-Taylor}. Moreover, oft-cited necessities for quantum gravity (e.g. the Big Bang singularity) can be averted by modifying the Einstein action such that gravity becomes weaker at short distances and small time scales ~\cite{Mazumdar}.  Thus it is crucial to test whether fundamentally  gravity is a ``quantum" entity. Proposed tests of this question have traditionally focussed on specific models, phenomenology and cosmological observations (e.g. \cite{Kiefer,phenomenology,Mazumdar2,Brukner,Retzker}) but are yet to provide conclusive evidence. More recently, the idea of laboratory probes, proposed originally by Feynman \cite{Feynman}, have started to take hold, but the existing ideas emphasize looking for stochastic fluctuations in the gravitational force \cite{Page-Geilker,Hendrik-Bassi, Hu-Anastopoulos}. Here we open up an alternative direction: probing the quantum coherent aspect of gravity, which seems to offer an unambiguous, and much more prominent, witness of quantum gravity than existing laboratory based proposals.
\begin{figure}[b]
    \includegraphics[width=0.71
\columnwidth]{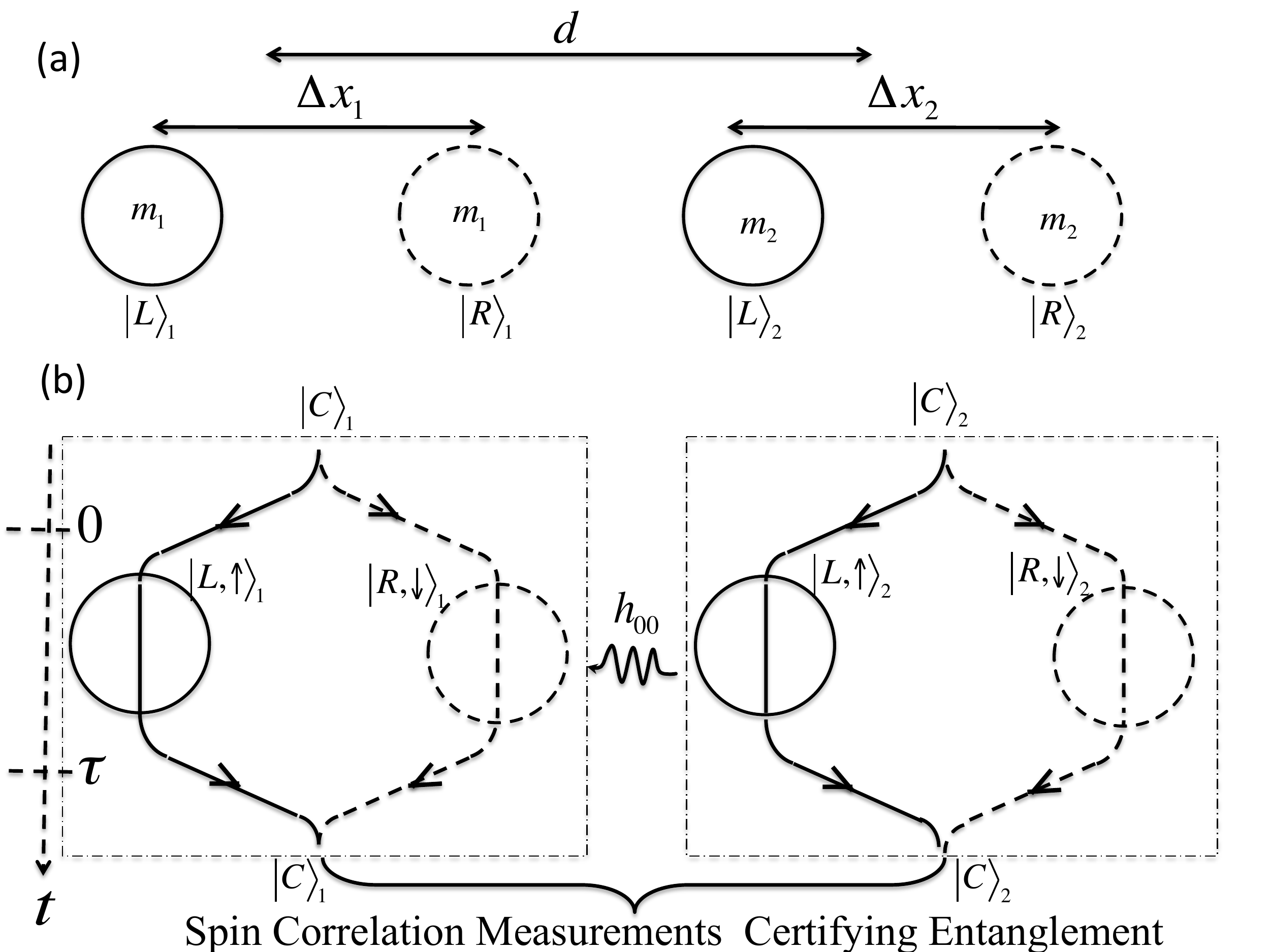}
 \caption{\footnotesize \textbf{Adjacent interferometers to test the quantum nature of gravity:} (a) Two test masses held adjacently in superposition of spatially localized states $|L\rangle$ and $|R\rangle$. (b) Adjacent Stern-Gerlach interferometers in which initial motional states $|C\rangle_j$ of masses are split in a spin dependent manner to prepare states $|L,\uparrow\rangle_j + |R,\downarrow\rangle_j$ ($j=1,2$). Evolution under mutual gravitational interaction for a time $\tau$ entangles the test masses by imparting appropriate phases to the components of the superposition. This entanglement can only result from the exchange of quantum mediators --  if all interactions aside gravity are absent, then this must be the gravitational field (labelled $h_{00}$ where $h_{\mu\nu}$ are weak perturbations on the flat space-time metric $\eta_{\mu\nu}$). This entanglement between test masses evidencing quantized gravity can be verified by completing each interferometer and measuring spin correlations.} \label{LRFig} 
\end{figure}
   
 We show that the growth of entanglement between two mesoscopic test masses in adjacent matter-wave interferometers (Fig.\ref{LRFig}(b)) can be used to certify the ``quantum" character of the mediator (gravitons) of the gravitational interaction -- in the same spirit as a Bell-inequality certifies  the ``non-local"  character of quantum mechanics and in the same vein that quantum entanglement can enable the probing of unknown processes \cite{Wehner,Paterek}. We make two striking observations that make the test for quantum gravity accessible with feasible advances in interferometry: (i) For mesoscopic test masses $\sim 10^{-14}$ kgs (with which intereference experiments might soon be possible \cite{Romero-Isart}) separated by $\sim 100~\mu$m, the quantum mechanical phase $\frac{E \tau}{\hbar}$ induced by their gravitational interaction (with $E$ being their gravitational interaction energy, and $\tau \sim 1$s their interaction time) is significant enough to generate an observable entanglement between the masses; (ii) If we use test masses with embedded spins and a Stern-Gerlach scheme \cite{Wan, Folman} to implement our interferometry, then, at the end of the interferometry, the gravitational interaction of the test masses actually entangles their spins which are readily measured in complementary bases (necessary in order to witness entanglement). To identify ``which" quantum aspect of  gravity is responsible for the aforementioned entanglement, we use the linearized quantized version of Einstein's theory of gravity \cite{Gupta} to find that off-diagonal terms in the coherent state basis of the gravitational field modes (a uniquely quantum attribute)  is necessary. In particular, if, after  including a quantized gravitational field in our calculations explicitly, we make it ``classical" -- i.e., suppress by hand the off-diagonal terms in the coherent state basis of the gravitational field modes, we automatically destroy the generation of entanglement between the masses.

Our proposal relies on a simple assumption:  the gravitational interaction between two masses is mediated by a gravitational field (in other words, it is {\em not} a direct interaction-at-a-distance). Once we make this assumption, we use a central principle of quantum information theory: entanglement between two systems {\em cannot} be created by Local Operatons and Classical Communication (LOCC) \cite{Horodecki}. Translating to our setting,  any external fields (including the gravitational fields of other masses) acting on the test masses can only make local operations (LO) on their states, while a classical gravitational field propagating between the masses can only give a classical communication (CC) channel between them. These LOCC processes cannot entangle the states of the masses. Thus it immediately follows that if the mutual gravitational interaction entangles the state of two masses, then the mediating gravitational field is necessarily quantum mechanical in nature. One physical way to understand this is a through comparison to trapped ion quantum computation, where to entangle two ions a quantum bus, namely a quantized vibrational mode,  is necessary -- a classical bus does not serve the purpose. Our scheme is thus an explicit realization of the general idea that the entanglement of two accessible quantum systems mediated by a third ``inaccessible" system, can be used to evidence the quantum character of the mediator \cite{Paterek}. It follows Feynman's prescription \cite{Feynman} that in a two mass scenario, one should analyse the ``channel provided by gravitational field itself via the quantum mechanical amplitudes" -- the inescapable conclusion of such a treatment is the creation of entanglement between masses.

{\em Entanglement due to gravitational interaction:-} We first consider a schematic version that clarifies how the states of two neutral test masses 1 and 2 (masses $m_1$ and $m_2$), each held steadily in a steady superposition of two spatially separated states $|L\rangle$ and $|R\rangle$ as shown in Fig.\ref{LRFig}(a) for a time $\tau$, get entangled. We will describe later how this will be achieved following the scheme shown in Fig.\ref{LRFig}(b). Imagine the centres of $|L\rangle$ and $|R\rangle$ to be separated by a distance $\Delta x$, while each of the states $|L\rangle$ and $|R\rangle$ are localized Gaussian wavepackets whose widths are much less than $\Delta x$ so that we can assume $\langle L|R\rangle=0$. There is a separation $d$ between the centres of the superpositions as shown in Fig.\ref{LRFig}(a) so that even for the closest approach of the masses ($d-\Delta x$), the short-range Casimir-Polder force is negligible. Distinct components of the superposition have distinct gravitational interaction energies as the masses are separated by different distances and thereby will have different rates of phase evolution. Under these circumstances, the time evolution of the joint state of the two masses is purely due to their mutual gravitational interaction, and given by
\begin{eqnarray}
|\Psi(t=0)\rangle_{12} &=&\frac{1}{\sqrt{2}}( |L\rangle_1+ |R\rangle_1)\frac{1}{\sqrt{2}}( |L\rangle_2+ |R\rangle_2)\nonumber\\\label{Initial}\\\rightarrow |\Psi(t=\tau)\rangle_{12} &=&\frac{e^{i\phi}}{\sqrt{2}}\{|L\rangle_1\frac{1}{\sqrt{2}}( |L\rangle_2+e^{i\Delta\phi_{LR}}|R\rangle_2)\nonumber\\&+&|R\rangle_1\frac{1}{\sqrt{2}}(e^{i\Delta\phi_{RL}} |L\rangle_2+ |R\rangle_2)\} \label{evolved}
\end{eqnarray}
where $\Delta\phi_{RL} =\phi_{RL} -\phi, \Delta\phi_{LR}=\phi_{LR}-\phi$, and
\begin{eqnarray}
 \phi_{RL} \sim  \frac{G m_1 m_2 \tau}{\hbar (d-\Delta x)},\,\,\,\phi_{LR} \sim  \frac{G m_1 m_2 \tau}{\hbar (d+\Delta x)},\,\,\, \phi \sim \frac{G m_1 m_2 \tau}{\hbar d}.
\nonumber
\end{eqnarray}
One can now think of each mass as an effective ``orbital qubit" with its two states being the spatial states $|L\rangle$ and $|R\rangle$, which we can call orbital states or spatial bins. Essentially we have dichotomized the motional state of the masses so that these can be treated as encoding a qubit in their spatial degree of freedom. As long as $\frac{1}{\sqrt{2}}( |L\rangle_2+e^{i\Delta\phi_{LR}}|R\rangle_2)$ and $\frac{1}{\sqrt{2}}(e^{i\Delta\phi_{RL}} |L\rangle_2+ |R\rangle_2)$ are not the {\em same} state (which is very generic, happens for any $\Delta \phi_{LR} + \Delta \phi_{RL} \neq 2n\pi$, with integral $n$), it is clear that the state $|\Psi(t=\tau)\rangle_{12}$ cannot be factorized and is thereby in an entangled state of the two orbital qubits. Witnessing this entanglement  between the spatial states of the masses then suffices to prove that a quantum field must have mediated the entanglement between them as a classical field {\em cannot} entangle the states of spatially separated objects.  In fact, the entanglement between the massive orbital qubits increases monotonically over $\Delta \phi_{LR} + \Delta \phi_{RL}$ evolving from $0$ to $\pi$, with the entanglement being maximal for $\pi$. Thus it is worth seeking conditions such that 
\begin{equation}
\label{cond}
\Delta \phi_{LR} + \Delta \phi_{RL} \sim O(1)
\end{equation}
so that the entanglement between the states of the masses is significant and can be easily witnessed.

It makes sense to start with particles of the largest possible masses, namely $m_1\sim m_2 \sim 10^{-14}$kgs for which there have already been realistic proposals for creating superpositions of spatially separated states such as $|L\rangle$ and $|R\rangle$ \cite{Romero-Isart}, although in this work, we will adopt (with some modifications) the mechanism of Ref. \cite{Wan}. Typically (such as for diamond) these are micro-crystals of radius $r\sim 1~\mu$m. Note that we are constrained to design an experiment in which only the gravitational interaction is active. This means that for $r\sim 1~\mu$m micro-spheres, the allowed distance of closest approach is $d-\Delta x \approx 200 \mu$m, which is the distance at which the Casimir-Polder interaction \cite{Casimir, Casimir-Polder} (estimation to be shown later) is a tenth of the gravitational interaction. The gravitational inverse
square law has been tested at the percent level at these
distances \cite{Newton}.
Now, just to get a basic estimate of the parameters which are needed to satisfy Eq.(\ref{cond}), let us consider the situation when the masses are separated by a far shorter distance for the $RL$ component of the superposition than for the other three configurations, i.e., 
\begin{equation}
d-\Delta x\ll d, \Delta x,
\nonumber
\end{equation}
we have
\begin{equation}
 \phi_{RL} \sim  \frac{G m_1 m_2 \tau}{\hbar (d-\Delta x)} \gg  \phi_{LR},  \phi.
\label{estimate-grav}
\end{equation}
Under these circumstances, we see that if the duration for which we can hold the superposition without decoherence is $\tau \sim 2$s then we can satisfy Eq.(\ref{cond}) and get a significant entanglement between the dichotomic spatial degrees of freedom of the two masses. In practice, it is very difficult to witness directly the entanglement between the dichotomized spatial orbital degrees of freedom as generated above as, for that, one will need to measure the spatial degrees of freedom in more than one spatial bases (which involves constructing ideal two port beam-splitters for massive objects). We next show how we naturally solve this problem by resorting to Stern-Gerlach (SG) interferometry which has recently been achieved with neutral atoms \cite{Folman}, and proposed for  freely propagating nano-crystals with embedded spins \cite{Wan}.

{\em Gravitational Entanglement Witnessing in SG Interferometry:-} The SG interferometer is not only a convenient means to create a superposition of spatially separated state; at the end of a SG interferometer, as we now show, the orbital entanglement is naturally mapped on to spin entanglement, and thereby much simpler to witness as spins are easily measurable in arbitrary bases. A SG interferometry (shown in Fig.\ref{LRFig}(b)) occurs in three steps, which we now illustrate:\\
\underline{Step 1}: A spin dependent spatial splitting of the centre of mass (COM) state of a test mass $m_j$ in the following way
\begin{equation}
|C\rangle_j \frac{1}{\sqrt{2}}(|\uparrow\rangle_j+|\downarrow\rangle_j) \rightarrow \frac{1}{\sqrt{2}}( |L,\uparrow\rangle_j+ |R,\downarrow\rangle_j),
\label{SGsup}
\end{equation}
where $|C\rangle$ is the initial localized state of $m_j$ at the centre of the axis of the SG apparatus and $|L\rangle$ and $|R\rangle$ are separated states localized on its opposite sides along the axis (these are qualitatively the same ones as shown in Fig.\ref{LRFig}). \\
\underline{Step 2}:  "Holding" the coherent superposition created above (Eq.(\ref{SGsup}) for a time $\tau$ (Consider the magnetic field of the SG effectively switched off for a duration $\tau$).\\
\underline{Step 3}: The third and final step brings back the superposition through the unitary transformations
\begin{equation}
|L,\uparrow\rangle_j \rightarrow |C,\uparrow\rangle_j, ~ |R,\downarrow\rangle_j \rightarrow |C,\downarrow\rangle_j,
\label{back}
\end{equation}
which is, essentially, a refocussing SG apparatus with magnetic field homogeneity oriented oppositely to the apparatus in step 1 (although, in practice, it is best to keep the same magnetic field inhomogeneity and simply flip the spin so as to reverse the SG effect of step 1). 

 Let us now assume that two such SG interferometers with neutral test masses $m_1$ and $m_2$ operate in close proximity (but masses do not come so close as to have a significant Casimir-Polder interaction) as depicted in Fig.\ref{LRFig}(b). Moreover, we assume temporarily that the evolution time in steps 1 and 3 (when the spin-dependent splitting and recombination takes place) is much smaller than the time needed for the accumulation of a non-negligible gravitational phase. Then during the step 2 of the SG interferometry, due to the mutual gravitational interaction, the joint state of the two test masses evolves exactly as in Eq.(\ref{Initial})-Eq.(\ref{evolved})  with the orbital qubit states $|L\rangle_j$ and $|R\rangle_j$ replaced by ``spin-orbital" qubit states $|L,\uparrow\rangle_j$ and $|R,\downarrow\rangle_j$. When we follow-up the evolution of Eq.(\ref{evolved}) of spin-orbital qubits with the step 3 of Eq.(\ref{back}), then we obtain the state at the end of the SG interferometry to be
\begin{eqnarray}
|\Psi(t=t_{\text{End}})\rangle_{12}& =&\frac{1}{\sqrt{2}}\{|\uparrow\rangle_1\frac{1}{\sqrt{2}}( |\uparrow\rangle_2+e^{i\Delta\phi_{LR}}|\downarrow\rangle_2)\nonumber\\&+&|\downarrow\rangle_1\frac{1}{\sqrt{2}}(e^{i\Delta\phi_{RL}} |\uparrow\rangle_2+ |\downarrow\rangle_2)\}|C\rangle_1|C\rangle_2,\nonumber
\end{eqnarray}
where the unimportant overall phase factor outside the state has been omitted. The above is manifestly an entangled state of the spins of the two test masses. It can be verified by measuring the spin correlations in two complementary bases in order to estimate the entanglement witness ${\cal W}=|\langle \sigma_x^{(1)} \otimes \sigma_z^{(2)} \rangle - \langle \sigma_y^{(1)} \otimes \sigma_z^{(2)} \rangle|$. If ${\cal W}$ is found to exceed unity then the state is proven to be entangled, and, thereby, the mediator, the gravitational field, a quantum entity.

  {\em An Explicit Scheme:-} We now outline an explicit interferometer. Each SG interferometer has to be fed in by neutral masses with a very low internal temperature and an embedded electronic spin. We can assume a scenario where they are released  simultaneously from two adjacent traps separated by $d\sim 450~\mu$m, and fall vertically through their respective interferometers as, for example, envisaged in Refs.\cite{Romero-Isart,Wan,Ulbricht}.  Micro-diamonds with an embedded NV centre spin is one candidate for being the test mass -- they can be trapped in diamagnetic traps \cite{colorado} and we have to aim to cryogenically cool their internal temperature to $\sim 0.15$ K. Alternatively objects such as Yb micro-crystals with a single doped atomic two-level system in optical traps can be cooled in their internal temperature by laser refirgeration. Any charges should be neutralized immediately following their release from their traps by demonstrated means \cite{neutralize}. The core aim is to drop two objects simultaneously -- one through each interferometer -- so that their states can become entangled through their mutual gravitational interaction while they traverse their respective interferometers. To this end, we adopt, in each interferometer, a modified version of the SG interferometry scheme of Ref.\cite{Wan} for splitting into two parts and then recombining the wavepacket of each mass in the horizontal direction while they fall vertically through the interferometer. In step 1 of the SG interferometer described schematically by Eq.(\ref{SGsup}), the test masses are subjected to an inhomogeneous magnetic field gradient in the horizontal direction for a time $\tau_{\text{acc}}$ with a spin-flip (by a short microwave $\pi/2$ pulse) exactly midway at time $\tau_{\text{acc}}/2$. Thus the initial state of each mass (say, a Gaussian wavepacket just below their respective trap location) is subjected to a spin dependent acceleration and decceleration in sequence, to reach at time $\tau_{\text{acc}}$ a superposition of spatially separated states $|L,\uparrow\rangle_j$ and $|R,\downarrow\rangle_j$ centred at $x_{j,L}$ and $x_{j,R}$ respectively with a spatial separation of
\begin{equation}
\Delta x = |x_{j,L}-x_{j,R}| \sim \frac{1}{2}\frac{g \mu_B \partial_x{B}}{m_j}  \tau_{\text{acc}}^2,
\end{equation}
where $\mu_B$ is the Bohr magneton, $g\sim 2$ the electronic $g$-factor and $\partial_x{B}$ the field gradient in the horizontal ($x$) direction. 
For a micro-object of mass $m\sim 10^{-14}$ kg, a magnetic field gradient of $\sim 10^6$ T m$^{-1}$ \cite{Wan} and a time $\tau_{\text{acc}}\sim 500$ ms, $\Delta x \sim 250 \mu$m. Note that this corresponds to the masses being separated by $d-\Delta x \sim 200 \mu$m at their closest approach at which the Casimir-Polder potential \cite{Casimir,Casimir-Polder} $\sim \frac{1}{(4\pi\epsilon_0)^2}\frac{23 \hbar c R^6}{4 \pi (d-\Delta x)^7} (\frac{\epsilon-1}{\epsilon+2})^2 \sim 0.1$ of the gravitational potential, where we have taken $R\sim 1 \mu$m as the radius of each micro-object (assumed spherical) and $\epsilon \sim 5.7$ as the dielectric constant of diamond. At this stage, step 2 is carried out: A microwave pulse is used to swap the electronic state to the nuclear spin state, so that the masses are not subjected to spin dependent forces any more, and evolve by falling in parallel next to each other for a time $\tau$. If we allow only a time of $\tau \sim 2.5$ s for this step, then the masses continue to fall parallel to each other to a very good approximation: their movement towards each other due to their gravitational acceleration towards each other $Gm/(d-\Delta x)^2 \sim 10^{-16}$ms$^{-2}$ is truly negligible.  Under these circumstances, given the different steady separations $|x_{1,\xi}-x_{2,\xi^{'}}|$ (where $\xi,\xi^{'}\in\{L,R\}$) the phases $\Delta\phi_{LR}\sim - 0.2$ and $\Delta\phi_{RL} \sim 0.7$ accumulated due to the gravitational interaction over the time $\tau\sim 2.5$s. This phase accumulation alone gives ${\cal W} \sim 1.16$ implying a gravitationally mediated spin entanglement. In practice, the witness will give a larger value as phase accumulation and the adjoining entanglement growth happens also during steps 1 and 3 of the SG interferometry.

{\em Decoherence:-} We require both the orbital and spin degree of freedom of the mases to remain coherent for the whole duration of the experiment. As we map to nuclear spins for step 2 of the interferometry with their very long coherence times, we only require electronic spins coherent for $1$s (in steps 1 and 3), which should be possible for micro-diamond below 77 K \cite{Bar-Gill} with dynamical decoupling pulses on its spin bath \cite{Kara}. To estimate collisional and thermal decoherence times~\cite{hornberger2004theory, romero2011quantum,carlesso2016decoherence} of the orbital degree of freedom we consider the pressure $P=10^{-15}$ Pa and the temperature $0.15$ K: the collisional decoherence time for a superposition size of $\Delta x \sim 250 \,\mu$m is the same order of magnitude as the total microsphere's fall time $\tau+2\tau_{\text{acc}}\sim 3.5$~s, while the thermal decoherence mechanism, due to scattering, emission and absorption of environmental photons, is negligible. Note that speculated spontaneous collapse mechanisms \cite{Penrose,Diosi,GRW-review}, if true, will typically lead to a strong loss of coherence on the time-scale of the experiment and inhibit the gravitationally mediated entanglement. A pivotally important stage preceding the entangling experiment is to take the interferometers far apart from each other to characterize the relative phases between the two paths in each SG interferometer as affected by nearby surfaces, other masses etc. While these are LO and thereby cannot give spurious entanglement between the test masses, the spin operators used in the Witness ${\cal W}$ will have to readjusted in accordance to these local phases. Note that although the internal cooling is necessary, the centre of mass motion of the test masses, if originally released from $\sim 1$ MHz traps, are allowed to have a temperature as high as $100$ K as that causes only a factor of $\sim 10^{-2}$ change in the gravitational phase, while the change due to spreading of the wavepacket during the experiment is truly negligible.

{\em Overcoming challenges:-} The significant value of ${\cal W}$ that we demand necessitates the long times 
 $\tau+2\tau_{\text{acc}}$ and high $\partial_x B$.  The latter could be achievable in principle using superconducting magnets with embedded flux \cite{Tomita} with the test mass staying $5 \mu$m from the surface of a $10 \mu$m magnet (will require motorized magnets co-moving with the masses maintaining constant distances from them; there will be four magnets, one close to each of the spin dependent position components of each mass), while the former could be met in a drop tower or space-based experiment. Smaller $\partial_x B$, enabling placements of magnets at a further/fixed distance from the masses, and shorter durations of the experiment permitting normal laboratories, are affordable if much smaller phases can be detected from the spin measurements to check $\Delta \phi_{RL}+\Delta \phi_{LR} \neq 2n\pi$. For example, a much more modest $\partial_x B \sim 10^4$ T m$^{-1}$ would prepare $\Delta x \sim 1 \mu$m superpositions and require us to detect $\Delta \phi_{RL}+\Delta \phi_{LR} \sim
\frac{G m_1 m_2 \tau}{\hbar d} (\frac{\Delta x}{d})^2 \sim 10^{-4}$. The low internal temperatures necessary for lowering the matter wave decoherence could be achieved in principle by liquid helium-3 droplets around the micro-object while in their traps \cite{Weilert}. Potential charge multipoles existing inside the objects could be averaged out  by setting the test masses spinning while they traverse the interferometer (recently this has been demonstrated with circularly polarized light \cite{Hoang}). If a  diamagnetic material such as diamond (magnetic susceptibility $\chi_m \sim 10^{-5}$) is used, one will also need to cancel the overall field on the microspheres (say, by an external field) so that the two masses do not interact directly magnetostatically via induced magnetic moments. As this interaction $\propto \chi_m^2$ a tailored material of $\chi_m \sim 10^{-7}$ will suffice to keep the magnetic field variation over the microsphere small enough to affect the above cancellation. Tailored materials, for example, coating a diamond by a much denser material, may also indeed be used to enhance the gravitational interaction with respect to the Casimir-Polder interaction.

 {\em Which quantum aspect of gravity is probed?} Let us now write down (somewhat schematically) the evolution of the system during step 2 of the SG interferometry including explicitly the quantum state of the gravitational field assuming the standard low energy effective field theory for gravity. This will explicitly show which qualitative aspect of the quantum nature of the field (namely the fact that it can stay in a superposition of distinct states) is {\em necessary} to entangle the masses. Here one quantizes small deviations $h_{\mu\nu}$ about the Minkowskian metric (formally $g_{\mu \nu}=\eta_{\mu \nu} + h_{\mu \nu}$ where $g_{\mu \nu}$ is the full metric, $\eta_{\mu \nu}$ the Minkowskian metric, and $h_{\mu \nu}$ small deviations on that). In the laboratory frame one can always define a unique time direction -- thereby we can unambiguously write a Hamiltonian. Moreoever, the velocities of the masses at any point of time are assumed to be much smaller than the speed of light $c$ so that only the $T_{00}$ component (the mass-energy density) of the energy-momentum tensor of the matter field is non-negligible, and thereby only the $h_{00}$ component evolves. Typically the mass $m_j$ would be the excitation of a continuum neutral scalar field, but in our case only two possible localized states are allowed in each interferometer among all the available spatial positions ${\bf r}$. We will thus schematically represent the state in each interferometer as a dichotomic state: $a^{\dagger}_{j,\xi}$ ($\xi \in\{L,R\}$) stand for the creation operators of a mass $m_j$ in a spatially localized wavepacket around the point ${\bf r}_{j,\xi}=x_{j,\xi}\hat{e}_x$ (as the motion in the $z$ direction does not change the mutual interaction during step 2 of the SG interferometer, we set $z=0$). The total Hamiltonian thereby is (assuming a volume $V\rightarrow \infty$ over which the gravitational field is normalized)
\begin{eqnarray}
{\cal H} &=& \sum_{j,\xi} m_j c^2 a^{\dagger}_{j,\xi} a_{j,\xi}+ \sum_{{\bf k},\lambda} \hbar \omega_k b^{\dagger}_{{\bf k},\lambda} b_{{\bf k},\lambda}\nonumber\\&-&\hbar \sum_{j,{\bf k},\lambda,\xi} g_{j,{\bf k}} a^{\dagger}_{j,\xi} a_{j,\xi} (b_{{\bf k},\lambda}e^{i{\bf k}.{\bf r}_{j,\xi}}+b^{\dagger}_{{\bf k},\lambda}e^{-i{\bf k}.{\bf r}_{j,\xi}}),
\end{eqnarray}
where $\omega_k=c k$, $k=|{\bf k}|$, $\lambda \in \{{\bf \times},{\bf +}\}$ are two independent polarizations of the gravitational field modes (creation operator $b^{\dagger}_{{\bf k},\lambda}$) and the last term (interaction term) originates from the interaction Hamiltonian \cite{Oniga} ${\cal H}_{\text{int},j} = -\frac{1}{2} \int h_{00} T_{00} d^3{\bf r}$ with coupling constants $g_{j,{\bf k}}= m_j c^2 \sqrt{\frac{2\pi G}{\hbar c^3 k V}}$. We have taken the view that the scalar fields representing the masses interact simultaneously with the common gravitational field modes, albeit at different locations, and ignored the propagation of the field from one mass to another as the masses are very close. Time evolution according to the above Hamiltonian has been exactly solved earlier in the context of quantum optomechanics \cite{Bose}, which we can readily apply to find the time evolution of the joint state of the matter and gravitational fields (for {\em arbitrary} initial coherent states $|\beta_{\bf k}\rangle_{{\bf k},\lambda}$ of the field modes)
\begin{eqnarray}
&&|\Psi(t)\rangle_{\text{mat+grav}} = \frac{1}{2} \sum_{\xi,\xi^{'}\in\{L,R\}} a^{\dagger}_{1,\xi} a^{\dagger}_{2,\xi^{'}} |0\rangle \nonumber\\&\otimes&
\prod_{{\bf k},\lambda} e^{i\frac{(g_{1,{\bf k}}e^{i{\bf k}.{\bf r}_{1,\xi}}+g_{2,{\bf k}}e^{i{\bf k}.{\bf r}_{2,\xi^{'}}})^2}{\omega_k}t}|\alpha_{{\bf k},\xi,\xi^{'}}\rangle_{{\bf k},\lambda}, 
\label{joint}
\end{eqnarray}
where $|\alpha_{{\bf k},\xi,\xi^{'}}\rangle_{{\bf k},\lambda}$ are {\em coherent states} of the gravitational field with
amplitude $\alpha_{{\bf k},\xi,\xi^{'}}=(\beta_{\bf k}+\frac{g_{1,{\bf k}}}{\omega_k}e^{i{\bf k}.{\bf r}_{1,\sigma,t}}+\frac{g_{2,{\bf k}}}{\omega_k}e^{i{\bf k}.{\bf r}_{2,\sigma^{'},t}})(e^{i\omega_k t}-1)$ (a small phase factor and a global phase factor has been omitted).

Note that the overlaps between coherent states $|\alpha_{{\bf k},\xi,\xi^{'}}\rangle_{{\bf k},\lambda}$ of a given mode ${\bf k},\lambda$ are exceedingly high so that the joint state can be well {\em approximated} at all times as a product state of the gravitational field and the masses so that the cross terms $\propto \frac{g_{1,{\bf k}} g_{2,{\bf k}}}{\omega_k} \propto \frac{1}{k^2}$ in the phase factor in Eq.(\ref{joint}) can be evaluated as an integral over momentum space to 
precisely give the gravitational mutual interaction induced phase shifts $\Delta \phi_{\xi,\xi^{'}}$. On the other hand, the factorization is only approximate and strictly the state in Eq.(\ref{joint}) is manifestly entangled (albeit an extremely low amount) with the gravitational field. If the field was artificially made classical by destroying all off diagonal terms $|\alpha_{{\bf k},\xi,\xi^{'}}\rangle\langle \alpha_{{\bf k},\xi^{''},\xi^{'''}}|_{{\bf k},\lambda}$, then the coherence between the distinct states of the matter field would also be instantaneously lost. Note that such a complete destruction of the off-diagonal terms amounts of reading reliably (or making a reliable record somewhere of) the highly non-orthogonal (nearly overlapping) states which should be impossible following unitary dynamics -- so this is indeed an artificial process just to ascertain that it is the quantum coherence between the coherent states $|\alpha_{{\bf k},\xi,\xi^{'}}\rangle_{{\bf k},\lambda}$ of the gravitational field that allow for the gravitationally mediated entanglement.

{\em Summary:-} How to consistently combine gravitational fields and quantized matter for a unified theory of physics has been an important issue. While gravity is one of the fundamental forces, its weakness has made it difficult to test theories on its nature. In particular, in order to treat gravity in the context of quantum mechanics, it is important to answer the question, ‘is gravity a quantum entity?’ Lack of a scheme to test this question has been a long-standing issue. In this paper, based on the definition of quantum correlations, we introduce an idea to solve this problem: to observe the entanglement of two test masses to ascertain whether the gravitational field is a quantum entity that follows the principle of quantum superposition. Instead of using the gravity of one test mass to change the position of another\cite{Page-Geilker, Hendrik-Bassi, Hu-Anastopoulos, Aspelmeyer, Mari}, which is a tiny effect to measure for a test mass as small as those for which large quantum superpositions are feasible, we consider a change of the phase affected by the gravitational interaction, which we find to be relatively large even for a pair of small test mass. Thus the test described here gives a much more prominent witness than tests of gravitational fluctuations, and is several orders of magnitude stronger than other predictions in the low-energy long-distance sector of quantum gravity such as post-Newtonian corrections \cite{Donoghue,Bohr} and decoherence induced by the gravitational field background \cite{Hu,Blencowe,Oniga}. The prescriptions we have provided for overcoming the challenges will set out a roadmap towards such experiments and could have other beneficial spin-offs on the way, such as the measurement of the Newtonian potential for microspheres, given that so far it has only been measured for much larger masses (this will only need one interferometer and a proximal mass) \cite{Aspelmeyer, Geraci, Newton}. Our proposal also includes a transfer of entanglement from phase space of the motion of the test mass to the state of spins embedded in the test masses. Thus the idea and scheme presented in this paper arguably opens up the shortest route known to date for a laboratory experiment of quantum gravity.

\textbf{Note Added:}  After this work was completed, we became aware of similar independent work by Marletto and Vedral (on arXiv today).

\textbf{Acknowledgements.} Significant progress on the work took place during the KIAS Workshop on ``Nonclassicalities in Macroscopic Systems" (August 2016). This work was presented in the ECT* workshop on ``Testing the limits of the quantum superposition principle", the Benasque Workshop on ``Quantum Engineering of Levitated Systems" (September 2016), the ICTS Bangalore Discussion Meeting on ``Fundamental Problems of Quantum Physics" (December 2016), and Quantum 2017, Torino (May, 2017), where feedback received from participants has been greatly beneficial. Particularly we would like to acknowledge incisive comments and questions by Miles Blencowe, Andrew Greentree, Jonathan Oppenheim, Anis Rahman, Lorenzo Maccone, Jack Harris and Gavin Brennen. HU and MT acknowledge funding by the Leverhulme Trust (RPG-2016-046) and the Foundational Questions Institute (FQXi). AG is supported by the U.S. National Science Foundation (PHY-1506431).
MSK acknowledges a Leverhulme Trust Research Grant [Project RPG-2014-055] and an EPSRC grant [EP/K034480/1]. MP acknowledges support from the EU Collaborative Project TherMiQ (grant agreement 618074), the SFI-DfE Investigator Programme (grant 15/IA/2864),
the Royal Society and the COST Action CA15220 “Quantum Technologies in Space”. GWM is supported by the Royal Society and the U.K. EPSRC “Networked Quantum Information Technology” Hub. (EP/M013243/1). PB and SB would like to acknowledge EPSRC grant EP/N031105/1. SB would like to acknowledge ERC grant PACOMANEDIA and EPSRC grant EP/K004077/1.




\end{document}